# Efficacy of mannan-oligosaccharide and live yeast feed additives on performance, rumen morphology, serum biochemical parameters and muscle morphometric characteristics in buffalo calves

Muhammad Zeeshan[1], Saima Masood[1*], Saima Ashraf[1], Shehla G. Bokhar[2], Hafsa Zainab[1], Saher Ijaz[1], Muhammad Usman[3], Ayesha Masood[4], Hafiz Faseeh U. Rehman[1], Mirza M. Usman[1]

[1]*Department of Anatomy and Histology, University of Animal and Veterinary Science- 54000 Lahore*

[2]*Department of Small Animal Clinical Sciences, University of Animal and Veterinary Science- 54000 Lahore*

[3]*Department of Basic Sciences (Histology), University College of Veterinary and Animal Sciences, Lahore (Narowal Campus) Narowal 51600, Pakistan*

[4]*University College of Veterinary and Animal Sciences, The Islamia University of Bahawalpur, 63000, Pakistan*

Corresponding Author: saima.masood@uvas.edu.pk

**Abstract**

The objective of the current study was to assess the effect of dietary supplementations of mannan-oligosaccharide, live yeast, and a combination of these two additives on growth performance, histo-morphology of the rumen, and muscle morphometric attributes in buffalo calves. A total of twenty buffalo calves (average weight of 25 ± 3kg) having 3 months of age were distributed according to a complete randomized design. All animals were individually stalled in the shed and were fed ad-libitum. Experimental animals were divided into four groups for 67 days: Control group(without the inclusion of dietary supplementation); MOS group (Mannan oligosaccharide 5 g/clave/day; Yeast group (Live yeast 2g/calve/day) and Mixed group (MOS + Live Yeast 2.5g + 1g )/calve/day. Experimental results revealed that combined supplementation of MOS and Yeast and MOS alone resulted in an increased number of short-chain fatty acids in the rumen as well as ruminal pH ($P<0.05$). Results showed a significant improvement in average daily gain and FCR of MOS and Mixed supplemented groups ($P<0.05$). Histomorphological evaluation of rumen mucosal epithelium showed a significant improvement in the mixed-supplemented group ($P<0.05$) as compared to the yeast-supplemented and control groups. Muscle quality parameters as meat texture showed significant improvement in MOS and mix-supplemented groups. Histological examination of longissimus dorsi muscle cross-section showed a significantly higher($P<0.05$) muscle fiber and muscle fascicle diameter in both MOS and mix-supplemented calves groups. In conclusion,the results of this experiment revealed thatdietary addition of MOS, Live yeast and their combination have positive effects on growth performance, rumen histology indices and muscle morphometric features in buffalo calves.

1. **Introduction**

Irrational use of antimicrobials in farm animals' health and production beyond therapeutic needs has been highlighted in recent years as one of the major risk factors for the acceleration of antimicrobial resistance of bacteria in both humans and animals (Haulisah *et al*., 2021). Besides treatment, a substantial amount of antibiotics is being used as feed additives to enhance the growth of animals. This overuse of antibiotics in livestock animals is considered one of the important factors that contribute to the dissemination and emergence of resistant bacteria (Chattopadhyay 2014; Truszczynski & Pejsak, 2006). Apart from the excessive use of antibiotics, the other main problem is the highly concentrated diets used to boost up livestock production. Though, this type of diet may severely affect the ruminal milieu due to the presence of rapidly fermentable carbohydrate following ruminal acidosis, hepatic abscess and foot problem. Therefore, ruminal acidosis of alarming disorder due to its difficult diagnosis in feedlot system which leads to the economic and health consequences (Diaz *et al*., 2018). To counter such issues, alternative strategies should be made for farm animal health and production and to cope with different infectious and metabolic disorders.

During the past decade, the scientist studied the use of natural products and organic minerals in monogastric as well as ruminants at critical stages like the adaptation period and transition period (Ferroni *et al*., 2020; Lopreiato *et al*., 2020). Pre- and pro-biotics recently gained attention due to their advantageous use as a dietary inclusion in animal feed such as complex carbohydrate molecules, mannan oligosaccharides (MOS), mannose known as mannoproteins, beta-glucans and proteins derived from the Saccharomyces cerevisiae yeast outer cell wall (Diaz *et al*., 2018). MOS provides competitive binding for gran-negative bacteria that have mannose-specific fimbriae type-1 helps in the attachment to D-mannose receptors of epithelium ultimately altering the composition of intestinal microbiota, digestibility and intestinal health of calves (McGuirk 2008).

The combination of pre-and pro-biotics in the animals is valuable in enhancing the animal immune response, growth and production subsequently opening new horizons in the constitution of these additives in animal feed. The possible outcome of these feed additives on the immune stimulation, microbiota digestive tract and pH regulation of the rumen are beneficial health, digestibility and performance of animals (Ovinge *et al*., 2018; Silberberg *et al*., 2013). Prebiotics could also be used to potentiate the effect of probiotics. The use of a combination of these products has synergistic effects which could be more efficient to stimulate intestinal micro-biota and also for the production of animals. But the selection of appropriate prebiotics and probiotic to generate a symbiotic formula is very important (Uyeno *et al*., 2015).

Probiotics are classified as beneficial microorganisms administered having immune-modulatory effect and are considered a good alternative to antibiotics. Different scientific studies provide confirmation on the beneficial effects of probiotics and prebiotics on the health of animals,especially in the meaning of protection against harmful pathogens, for the stimulation of immune response and to increase production capacity (Al-Shawi *et al*., 2020; Markowiak & Ślizewska, 2018). Albeit, some studies are available that describe the combined synergistic effect of MOS and live yeast in ruminants (small and large) production but certainly there is a dearth of literature that describe the effect of MOS and live yeast on the ruminal growth in terms of its cellular changes in calves. Therefore, this study was designed to evaluate the effect of these pre- and probiotics on the ruminal histomorphometry and meat characteristics in the calves.

## 2. Materials and methods

### 2.1 Preparation of Calves

This trial was performed at UVAS- Pattoki research farm, Lahore. A total number of 20 Nili Ravi buffalo calves of three months of age were housed separately with separate pens, each of which was equipped with feeding and watering according to the requirement of calves. All pens were located in the same calf house and randomly allocated. The calf house was equipped with a controlled ventilation system and the bedding With daily removal of manure. Temperature and air humidity were monitored.

### 2.2 Feeding Strategy of Animals

A commercially available calf starter in powder form was given to the calves. The diet consisting of calf starter and Total mixed ration (TMR) (Calf Starter + Hay + Molasses) was offered twice a day, early morning at (6:00 am) and evening at (6:00 pm). Prebiotic (Mannaoligosaccharide), Probiotic(Yeast culture Multi strain) and the combination were mixed in calf starter for 63 days of the experiment once a day early in the morning (6.00 am).

### 2.3 Experimental Design

The experimental animals were divided into four different groups (n=5). Different dietary treatments were given to these groups as follows
Control Group (T1): fed with only basal diet
Group 2 (T2): Provided with a basal diet supplemented with MOS (Mannan-Oligosaccharide) at a dose rate @ 5g/calf/day
Group 3 (T3): Provided with a basal diet supplemented with Probiotic (Yeast Culture) at a dose rate @2.5g/calf/day
Group 4 (T4): Combined group was provided basal diet + Prebiotic + Probioticat dose rate2.5g/calf/day respectively.
This treatment was done for a period of 9 weeks.

### 2.4 Growth Parameters

Body weight was measured on the first day of every week before the feeding early in the morning. Daily feed intake of the calves with calve starter and feed additives was recorded to calculate the feed conversion ratio (FCR). The weight of animals was checked and recorded in the start of the trial and during the trial to check weight gain on a fortnight basis then compare the readings to evaluate the weight gain by the supplementation of mannan oligosaccharides, antibiotics, and a combination of these additives.

## 2.5 Serum Analysis of Supplemented Calves

Blood was collected at the end of the trial before morning feeding. Ten ml blood sample was drawn by jugular puncture in non-EDTA vacutainer followed by centrifugation for 5 minutes at 3900 rpm. The serum was collected in a 2ml Eppendorf tube using a capillary pipette and stored at a chilled temperature for the study of biochemical parameters.The serum biochemistry like Glucose, Cholesterol, triglyceroids, MDA and calatasle were determined. Apart from them, liver function test including AST and ALT wew also determined. For serum cholesterol determination blood chemistry analysis was done.

## 2.6 Sample collection

Meat samples were obtained after the slaughtering of the calves. Meat samples were obtained from the Longissimus dorsi muscle for histomorphometry, pH and water-holding capacity (WHC). Similarly, for ruminal and small intestinal samples, the digest was removed and washed with distilled water.A piece of about 1cm$^2$ was collected from the dorsal and ventral sac region of the rumen and from all three parts of the small intestine followed by washing in (0.1 molar, pH 7.4) phosphate buffer solution and fixation in neutral buffered formalin

## 2.7 Muscle Parameters

### 2.7.1 Measurement of Water Holding Capacity (WHC)

Honikel's gravimetric drip loss method was used as described by Honikel (1998). Muscle samples were weighed initially placed in a meshed pouch and suspended in a special container having lid to stop evaporation and were placed in a refrigerator at 4-6 °C for 48 hours.The direction of muscle fiber in the samples should be horizontal to gravity (Rasmussen and Andersson 1996).After 48 hours samples were removed from the refrigerator and weighed again. The percent difference in the two weights was used for calculating water holding capacity. Formula used for WHC calculation is

$$WHC = (W1-W2/W1)*100$$

### 2.7.2 Measurement of Muscle pH

The muscle pH of Longissimus dorsi meat samples was measured with a digital pH meter. A vertical deep incision was given in the muscles and after calibration of pH meter piercing Knob of a digital pH meter was inserted 1 cm deep into the muscles and pH was measured within 15 minutes after slaughtering. Muscle pH was determined at time intervals of slaughtering time, 12 and 24 hours respectively.

### 2.7.3 Shear force value

Shear force tenderness of the meat.Tendernessisconsideredimportantforthe determination of the quality of meat. Tenderness is measured by using the method of slices (Aberle and Forrest 2001) To calculate the shear force value strips of 1cm$^2$ was made and a micro stable system (Texture Analyzer) was used to check the shear force. Boiled meat was used for this purpose.

2.8 Histological process

The fixed samples were subjected to grading concentrations of ethanol for dehydration following clearing and clearing and infiltration with paraffin. The paraffine blocks were prepared on the embedding desk and sections of 5µm thickness were cut through the microtome. These sections were mounted on slides and then stained with hematoxylin and eosin techniques (Svarna *et al*., 2019). All histological slides were observed through the light microscope (LABOMED®USA) connected through a camera and computer which is coupled with a computer by using Pixel Pro imaging software 2.0.

2.9 Histomorphometrical analysis

The histomorphometry of rumen and muscles were observed through the software (Prog Res® 2.1.1 Capture Prog Camera Control Software). All observations were made on three tissue sections of each animal. Tissue sections of rumen were analyzed for the measurement of epithelium thickness, and connective tissue thickness width., while for muscles, muscle fascicle and fiber diameter along with muscle fiber density were estimated. For muscle per unit area count, circle of 0.5mm was made by using mentioned software to mark the area. Then this circle is divided in two halves. All the muscle fibers located in the circle along with muscle fibers touching the boundary on the right half side of the circle were counted. Muscle fibers coming under the boundary line of the left side were excluded from this counting.

2.10 Statistical Analysis

The distribution of data was determined by Kolmogorov Smirnov test. One way analysis of variance (ANOVA) was used for data analysis by using Statistical Packages for Social Sciences data presented as mean ±D. Duncan's Multiple Range tests was used to compare group differences. P<0.05 was considered significant.

3. Results

3.1 Growth Parameters

3.1.1 Determination of Feed Conversion Ratio (FCR)

The least value of FCR (6.75) was found in the symbiotic-supplemented group (T4) which is the most significant among the other three groups. However, the values of prebiotic and probiotic-supplemented groups were also significantly improved as compared to the control group.

3.1.2 Weight Gain

The pattern of weight gain during this experiment is represented in the figure 1. All the treated groups showed an increasing trend but the most significant increase was observed in T3 and T4, especially in 5$^{th}$, 7$^{th}$ and 9$^{th}$ week of the experiment. The weight gain of T4 during the last two weeks (9$^{th}$-week reading) is 19.02±2.73 kg which is the highest amongst the group followed by 17.12±1.98 and 15.23±2.8 of T3 and T2, respectively.

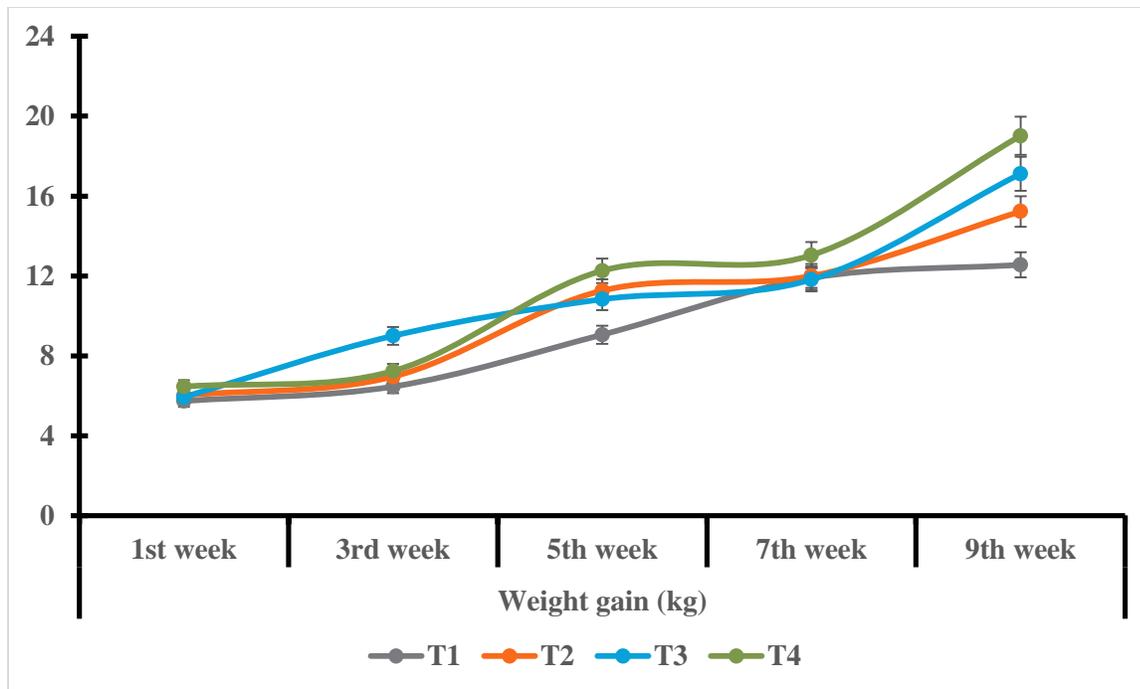

**Fig 1**: weight gain of buffalo calves treated with different feed additives during the experiment.

3.2 Shear Force Analysis

The values of symbiotic supplemented group were especially more significant than the other three groups, while the values of the prebiotic and probiotic-supplemented groups were also significantly improved as compared to the control group.

3.3 Determination of pH

P-Value is 0.058 After the slaughtering of animals, the pH of their carcass was taken. Range of pH values was from 5.5±0.5to6.0±0.3.Values of pH are within the range which is considered best for the quality of meat**.**

3.3 Serum Parameters:

The serum parameters are represented in figure 2.
The levels of liver markers namely AST and ALT were found within the normal limit however, the addition of MOS and yeast as symbionts in the T4 group significantly affect these values in the form of decreasing pattern as compared to other groups. Albeit, Glucose and urea levels did not show any significant changes with the treatments of MOS and yeast, but the level of triglycerides showed decreasing pattern with the most marked decrease in the T4 group. The same trend was followed by the MDA level whilst the catalase level exhibited the otherwise pattern with a significant increase in the T4 group amongst all other groups which didn't show any deviation.

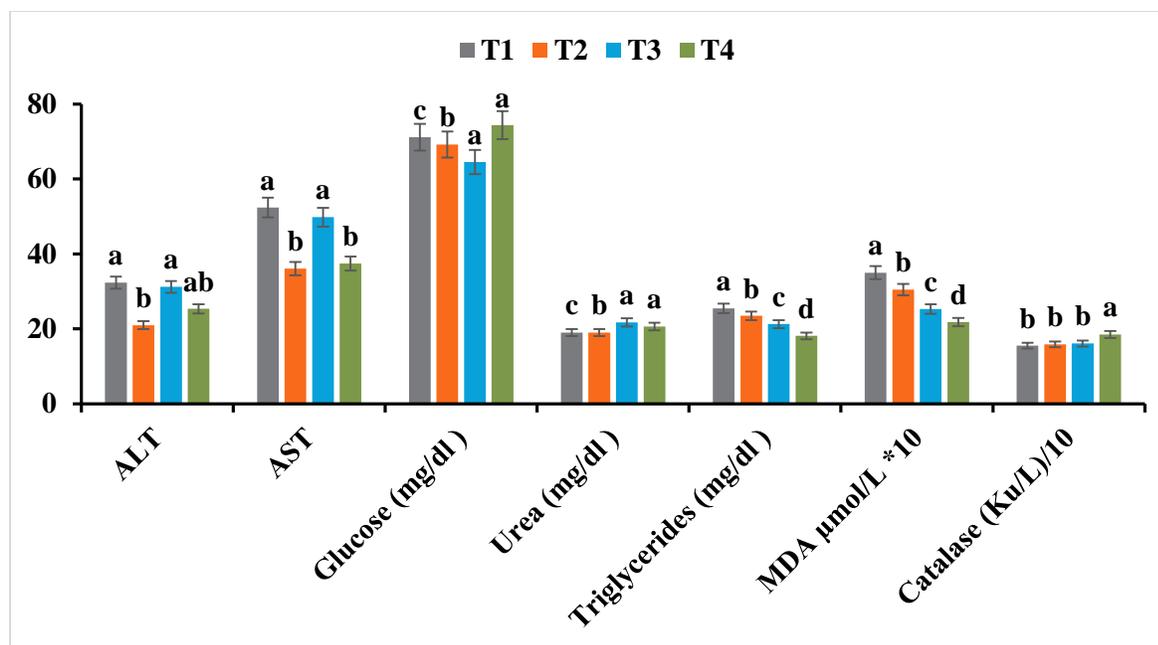

**Fig 2: Columns sharing different superscripts are statistically different at P<0.05**

3.4 Histomorphometry of Rumen and muscle

There was a significant effect as (p<0.05) in the papillae height (PH) and papillae width (PW) in the supplemented groups as compared to the control group as described in Table 1 and showed in figure 3. There was a significant improvement in the PH and PW of the rumen in the T4 (synbiotic-supplemented) and T2 (prebiotic supplemented group) while no significant improvements were observed in the control group and T3 (probiotic-supplemented group). The values of PH and PW for group T4 are 1957.84 ±24.23 and 117.18 ± 1.23μm, respectively, which is the highest as compared to other groups. Our results showed significant improvement in muscle fiber and muscle fascicle diameter of supplemented animals Table 1 and figure 4.

**Table 1**: Mean ± SEM values of different groups treated with prebiotics and Probiotics

| Parameters | Control (T1) | Prebiotic (T2) | Probiotic (T3) | Synbiotic (T4) | P-Value |
|---|---|---|---|---|---|
| **Histology of Ruminal Papillae** | | | | | |
| PH(μm) | 1758.57 ± 32.39 | 1831.49 ± 38.06 | 1756.10 ± 21.49 | 1957.84 ±24.23 | .001 |
| PW(μm) | 111.46 ± 0.78 | 114.3 ± 1.05 | 111.52 ± 1.39 | 117.18 ± 1.23 | .008 |
| **Histology of Muscle Tissue** | | | | | |
| MFb Di (μm) | 33.82 ± 1.13 | 35.00 ±0.87 | 33 ± 1.05 | 37.04 ± 0.66 | 0.044 |
| MFc Di (mm) | 0.336 ± 0.018 | 0.442 ± 0.008 | 0.360 ± 0.014 | 0.494 ± 0.006 | 0.000 |

Means are statistically different at P<0.05 in a row

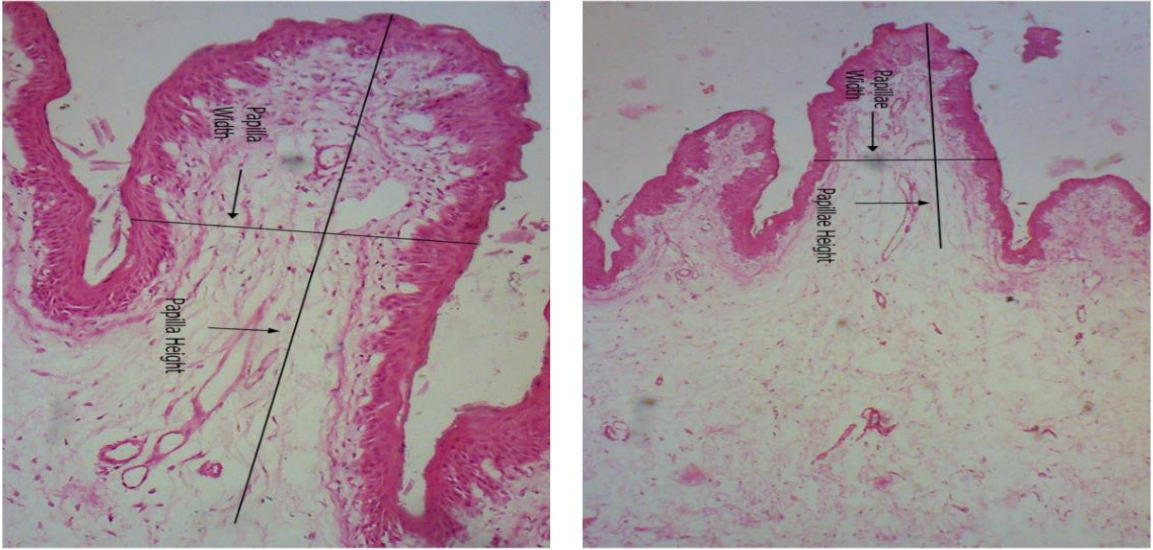

**Fig 3. Height and Width of Ruminal Papillae at 10x**

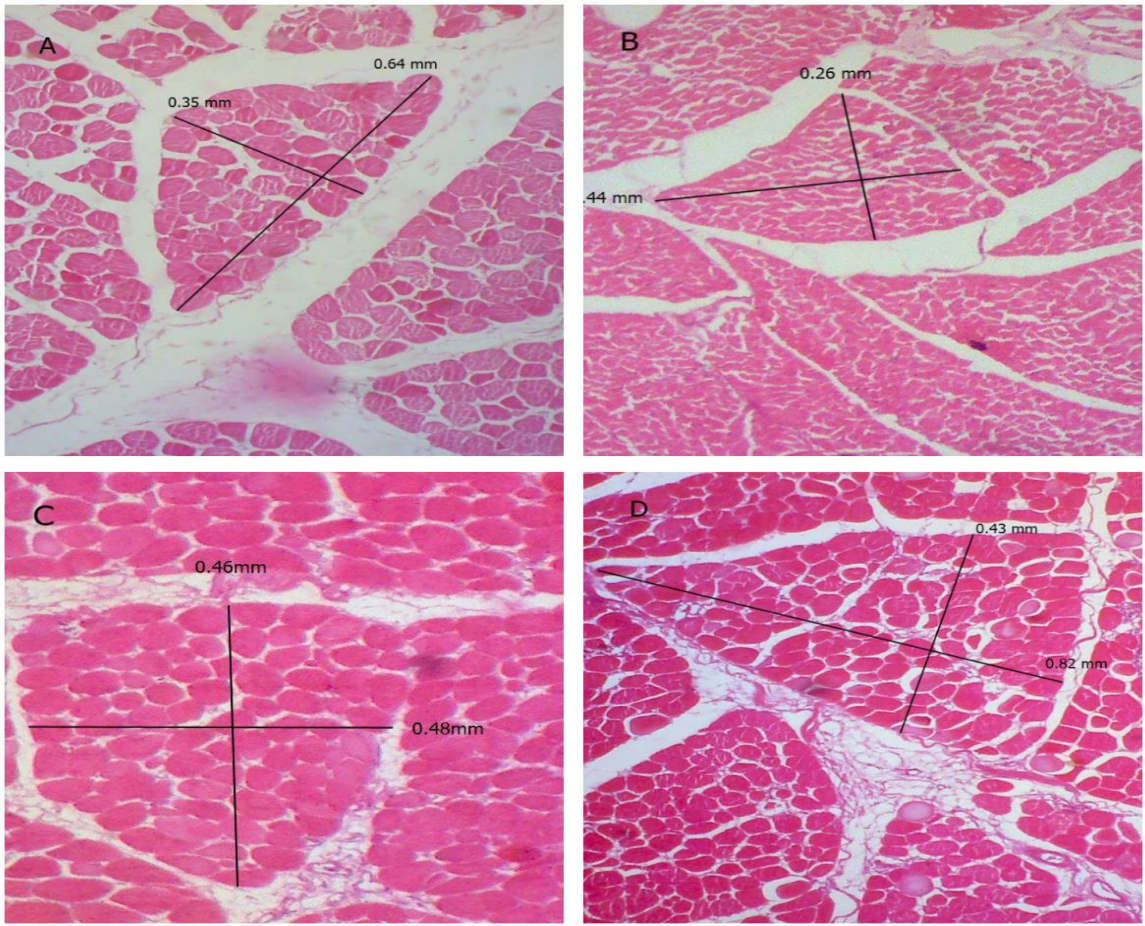

**Fig 4. Muscle Fascicle Diameter of supplemented Groups at 4x**

## 4. Discussion

Excessive and irrational use of antibiotics use as feed additives in livestock feed may alter the ruminal mallei and accelerate the antimicrobial resistance of bacteria. Therefore, recent era scientist urges the use of natural products to overcome this main issue. In this regard, pre- and probiotics employment has gained attention, hence this study was designed and conducted. The primary focus of this study is to evaluate the supplementation of prebiotic mannan oligosaccharide (MOS), and probiotic (yeast culture) effects in singular as well as in combination form on the ruminal histomorphometry, meat characteristics and serum analysis in calves. During the 9 weeks of trial, the growth performance in terms of FCR and weight gain was positively influenced by the MOS and yeast supplementation which is in line with the with results reported by (Maamouri & Ben Salem, 2021; Sharma *et al*., 2018; Zheng *et al*., 2021). But in these researches, a combination of yeast and MOS did not evaluate but rather separately supplemented.

Many scientific reports are available that describe the effect of MOS and yeast separately in ruminants. There is a paucity of data that elaborates the use of MOS and yeast combination in ruminants' calves. It is clear from these research efforts that yeast culture supplements can beneficially modify microbial activities, fermentation, and digestive functions in the rumen. Research has shown that viable yeast culture preparations can stimulate specific groups of beneficial bacteria in the rumen, and provide mechanisms that may explain their effects on animal performance (Uyeno *et al*., 2015). Due to excellent bacterium-biding properties (Tóth *et al*., 2020). MOS prevents the adherence of pathogenic bacteria to the enteric mucosa leading to improve FCR and Feed intake as witnessed in groups 2 & 3 but more significant improvement is seen in groups 4 or T4. These results are in line with the findings of (Grossi *et al*., 2021) in buffalo calves supplemented with a combination of MOS, yeast and selenium. Better health status (daily weight gain, FCR feed intake) of the calves can be attributed to the symbiotic property of MOS and yeast that apprehended the free radical production through better glutathione activity, thereby improving the immune function of animals (Alagawany *et al*., 2021; Sanchez *et al*., 2021).

Glucose level remained unaffected by the MOS and yeast supplementation separately as well as in combination. The same pattern of glucose concentration was reported by (Dar *et al*., 2019). However, (Raza *et al*., 2022) found that the glucose concentration was decreased in calves supplemented with yeast and MOS separately. The discrepancies in this result may be linked to anatomical and physiological changes in rumen development. During ruminal development in calves, The primary source of energy shifts from glucose to volatile fatty acids may overshadow the glucose concentration. The decreased trend was seen in the triglycerides level while urea level did now any remarkable change in the current study finding. This pattern is in accord with the reports of (Dar *et al*., 2019) and (Al-Saiady, 2010). The possible reasons for lower serum triglyceride levels might be either due to a decrease in intestinal lipid absorption or due to increased lipid catabolism. The liver function markers/enzymes, AST and ALT, represent the integrity of the liver function. Any change in these enzymes, either increase or decrease, indicates a faulty in the hepatic tissue architecture. The supplementation of MOS and yeast did not affect the levels of these enzymes which are in the agreements of (Diaz *et al*., 2018; Rekiel *et al*., 2007)). Their supplementation in livestock feed during stress condition reported to improved the liver function that likely due to immuno-modulatory and anti-inflammatory potential of the MOS and yeast in combination.

There was a significant improvement in the height and width of the ruminal papillae. As value of p<0.05which shows the significant improvement in the papillae height and width of treated animal groups. Our results were similar to the (Diaz *et al*., 2018) who found that by the use of prebiotics and probiotics as additives improvement in the health and size of ruminal papillae occurs. While in the intestinal region prebiotics helped to improve the micro-biota by decreasing the number of pathogenic bacteria and creating a friendly environment for the beneficial bacteria. Hence, it played a vital role in the improvement of the intestinal environment.There was a significant improvement in the height and width of ruminal papillae in the synbiotics-supplemented animals and prebiotic-supplemented group while there was no special improvements observed in the control group. With the increasee of papillae size and improvement of blood circulation, its absorption ability increased which played a vital role in the improvement of animal health.This combination improves the survival of organisms and provides a specific substrate for its fermentation (Collins & Gibson 1999). ShearForcevalueswere analyzed using the Texture Analyzed and the muscular shear force values were in the excellent category of meat evaluation parameters. The findings of the current study about the histomorphometry of muscles is similar to (Shah *et al*., 2020) and (Liu *et al*., 2023) who reported significant improvement in muscle fiber and muscle fascicle diameter of supplemented animals. While there was no significant improvement was observed in the control group. Albrecht *et al*. (2006) experimented on different cattle breeds to study the structure of muscle, muscle fiber diameter and muscle fascicle diameter which provided us comparative values to check our work. Similarly (Kiran *et al*., 2015) also experimented to study the ultra-muscle structure of buffalo meat and provided the parameters to compare muscle fiber and muscle fascicle diameter. Probiotics containing foods are a great hope of food industry as consumers don't like to accept products coming fromanted bacterial supplemented animals. Use of such feed additives like prebiotics and probiotics is safe and don't have any bad impact on health and environment.

## 4.1 Conclusions

Supplementation of prebiotics, probiotics and their combination improves the growth performance, Muscle Morphology, ruminal histology, intestinal microarchitecture and serum biochemical profile in buffalo calves.

## 5. Acknowledgments

The project is funded by Higher education Commission of Pakistan.